\documentstyle[12pt,titlepage]{article}

\renewcommand{\theequation}{\thesection.\arabic{equation}}

\outer\def\beginsection#1\par{\medbreak\bigskip
      \message{#1}\leftline{\bf#1}\nobreak\medskip
\vskip-\parskip
      \noindent}

\def\laq{\raise 0.4ex\hbox{$<$}\kern -0.8em\lower 0.62
ex\hbox{$\sim$}}
\def\gaq{\raise 0.4ex\hbox{$>$}\kern -0.7em\lower 0.62
ex\hbox{$\sim$}}

\def\beq{\begin{equation}}
\def\eeq{\end{equation}}
\def\bea{\begin{eqnarray}}
\def\eea{\end{eqnarray}}

\def \a {\alpha}

\begin{document}
\bibliographystyle {unsrt}

\begin{flushright}
CERN-TH/96-355
\end{flushright}
\begin{center}
SUMMARY AND OUTLOOK\\
G. Veneziano \\
Theoretical Physics Division, CERN \\
CH - 1211 Geneva 23
\end{center}

\vspace{10mm}

\renewcommand{\theequation}{1.\arabic{equation}}
\setcounter{equation}{0}
\section {Preamble}

A little over a year ago I gave the {\it summary} talk at SUSY-95 in
Paris.
It was quite a dreadful experience. I came to the Conference with
nothing in my
hands, sat through all the talks but one, preparing my transparencies
at
night,  skipping all social events. Very tired, I ended trying to
summarize
every talk, even the one which, not only I had missed, but also, as I
learned later,
had been cancelled\dots
This is why, this time, I decided for a compromise, taking advantage
of the fact
that I was also asked to provide an {\it outlook}. I came with a
number of prepared
transparencies, yet sat through all  plenary talks so that I could
add something
``on-line". The result is for you to judge: probably, the outlook
side of the story
somewhat overshadows the summary side: I apologize in advance
for having skipped over many important new results presented at the
Conference,
in particular over those on new detectors\cite{detectors} and on future
accelerators\cite{accelerators}, subjects which are far from my own
expertise.
Fortunately, you will find all the missing stuff in the preceding 27
plenary talks!
\vspace{2mm}

We are approaching the end of the century and we may ask ourselves:
What can we
learn from this century's experience in our field? It looks to me that
a
combination of
\begin{itemize}
\item new, solid experimental facts,
\item sound,  inspired theoretical thinking,
\end{itemize}
have paved the way to progress over and over again:
\begin{itemize}
\item The finite, constant speed of light led to Special Relativity via
the
equivalence of different Lorentz frames.
 \item The universality of free-fall gave
 General Relativity out of Einstein's generalized equivalence
principle.
\item The stability of atoms led to Quantum Mechanics through the
uncertainty
principle.
\item The wealth of particle data we accumulated for many decades, and
a
theoretically sound way to combine special
relativity and quantum mechanics, led to the Standard Model.
 \end{itemize}

The outlook side of my talk should try to answer the question:
What's next? You will see below how much/little I have dared to enter
suchslippery grounds.

\renewcommand{\theequation}{2.\arabic{equation}}
\setcounter{equation}{0}
\section {Listening to Nature}

One does not have to be very biased to conclude
that:
\begin{itemize}
\item Nature likes {\it local} symmetries.
\item At present, {\it all} phenomena
are well described by a {\it gauge} theory based on the group:
\beq
G_{gauge} = SU(3)_c \otimes SU(2)_L \otimes U(1)_Y \otimes Diff_4 \; ,
\label{G}
\eeq
 where the last, perhaps less familiar, factor stands for the
invariance group of
General Relativity (differentiable, general coordinate
transformations).
The gauge
bosons of $G_{gauge}$ (a photon, a graviton, eight gluons,
three intermediate vector
bosons) mediate the four known fundamental forces with couplings:
\beq
      \a, G_N,  \a_{s} , G_F \; .
\eeq
\item Although mass terms for the gauge bosons are forbidden,
long-range
gauge forces can be avoided  either through the Higgs mechanism or
through
confinement.
 We  believe that:
\begin{itemize}
\item Strong interactions  are realized (at zero temperature) in the
confining phase.
\item Weak interactions are realized (again at zero
temperature) in the Higgs phase.
\item Electromagnetic and gravitational interactions are realized in
the
Coulomb phase (i.e. with massless gauge bosons).
\end{itemize}
 As a result, two of the fundamental interactions are short range,
two are long
range.
\item The basic (left-handed spin $1/2$) constituents of matter belong
to a baroque (read highly reducible), {\it completely chiral}
representation $R$ of
$G_{gauge}$:
\bea
\Psi_L \in R &=& 3 \cdot [ (3,2,1/3,1/2) +
(\bar{3},1,-4/3,1/2) +
(\bar{3},1,2/3,1/2) +\nonumber \\
&& + (1,2,-1,1/2) + (1,1,2,1/2) ]\; ,
\label{R}
\eea
where the numbers inside the brackets indicate the representations
for the various factors appearing in (\ref{G}) and the overall factor
$3$ stands
for  family repetition. The corresponding (right-handed) antiparticles
 automatically transform
in the complex-conjugate representation $\bar R$.

 The physical meaning of a completely
chiral fermionic representation is that no
(gauge-invariant) mass term is allowed. Chiral fermions are thus
naturally light, i.e. their mass can only emerge from the (spontaneous)
breaking of the gauge symmetry.
\item There is, {\it probably},
an elementary scalar system, the complex Higgs-boson doublet $H$,
transforming as
\beq
H \in (1,2,1,0) + {\rm c.\; c.}
\eeq
This remains, to this day, the most uncertain element of the Standard
Model.

\end{itemize}

\renewcommand{\theequation}{3.\arabic{equation}}
\setcounter{equation}{0}
\section {The Standard Model at work}

Before turning my attention to the experimental status of the
Standard Model (SM),
I would like to make a short digression.
Today, more than ever, a sound equilibrium between theory and
experiments appears
to be crucial for our field to thrive. Precise data often fail to
convey an
 exciting message, either because we lack any idea about the
origin of a particular precisely measured number (e.g. the
fine-structure constant),
or because it just is very hard  to compute such number with
comparable precision
(e.g. the proton mass). Similarly, a grandiose theory making
untestable predictions (string theory?) remains forever in the realm of
pure
speculation.

{}From this point of view, the electroweak theory stands out, in the
sense that theoretical and experimental precisions appear to match
across the
board. If anything is lacking there, it is  a sign of  discrepancy
between the two, which makes the game somewhat boring.
At the other extreme we have gravity, endowed with a beautiful (though
only
classical) theoretical framework, General Relativity (GR),  whose
predicted
deviations from Newtonian gravity are often too tiny to be measured
(e.g.
gravitational waves). Things, however, are improving fast in that
area, as we
shall discuss below. Finally, with the advent of QCD, strong
interactions have seen
an enormous improvement, with theoretical predictions
often matching experimental precision in the case of hard processes.
For soft
physics, however, strong interaction theory is still lagging much
behind
experiments.

 How does the SM work in detail? This was the subject of many parallel
and plenary sessions at this Conference. I counted eight plenary
sessions on $SU(3)$, five on $SU(2)\otimes U(1)$, and one on GR,
roughlyin proportion to the number of generators$\dots$ I will limit
myself to highlight
what, in my opinion, was qualitatively new with respect, say, to a year
ago.

 \subsection {Strong interactions}
 As nicely expressed by R. Brock, a common trend in this area has been:
{\it ``An increased awareness of the glue"},
something indeed reflected in many of the topics discussed below.

\paragraph{\it 3.1.1~Hadron spectroscopy.}  My reaction to what I
heard on this topic was:
 Glueballs at last? For many years,
theorists have claimed that glueballs should exist, albeit within
admixtures with
 $q\bar{q}$ states.  It now looks\cite{Landua} that a few doubtful
previous
  candidates have magically turned into a single
convincing one.
 We can write this as an equation:
 \beq
G(1590) = f_0(1450) = f_0(1500) = \rm{glueball\;?}\; ,
\label{Glue}
\eeq
where the three above gluon candidates
have been reported, respectively, by GAMS, WA91 and  the Crystal Barrel
experiment. It seems that the apparent mass differences among these
states
can be explained in terms of experimental cuts and/or different
acceptances so that
the three  could very well be one and the same particle. But why a
glueball?

 Landua\cite{Landua} has discussed four glueball tests, which are all
passed by a
$O^{++}$ glueball. I can add  a fifth one, which actually goes back
to an old
paper\cite{NV} whose title is almost the same as eq.
(\ref{Glue}). Unlike what one would naively expect, a relatively pure
glueball
should decay more often
into $\eta(\eta')\eta(\eta')$ than into $2\pi$ or $K\bar{K}$, up to
phase-space
effects.  This is because the flavour-singlet pseudoscalar (a known
mixture
of $\eta\eta'$) gets its mass from the Adler-Bell-Jackiw (ABJ) anomaly,
a
phenomenon related to a purely gluonic channel. The
$\sigma\sigma\rightarrow4\pi$
channel should also be favoured, if the $\sigma$ (the broad $\pi\pi$
structure
around $700$ MeV) is somewhat mixed with the lightest scalar
glueball. All these expectations seem to be
consistent with the  data, suggesting that a positive answer to
eq. (\ref{Glue}) might  finally be in sight.

\paragraph{\it 3.1.2~Spin structure of the nucleon.}
The so-called spin crisis, rather than a real crisis of QCD, is yet
another
manifestation of the importance of the glue, being directly related to
the
just-mentioned ABJ anomaly. The latter, besides boosting up the
$\eta'$ mass (thus
solving the famous $U(1)$ problem), should also suppress the
flavour-singlet axial
current (which is what the spin crisis is  all about). On the
experimental front,
nice progress was reported\cite{Nass} by the CERN and SLAC
collaborations on the measurements of $g_1^{p,d,^3He}$ down to small
$x$. Together
with planned experiments on heavy-flavour production and on the
structure of the
hadronic final state, this should soon help  clarify the issues.

\paragraph{\it 3.1.3~Small-$x$ physics, soft interactions,
diffraction.}

We may call what has been happening in this area\cite{Pom} the
Pomeron's comeback.
It reminds me of Sid Coleman teasing some of us in the late 60's:
 ``At Harvard, we think
of the Pomeron as a French wine $\dots$" (for non-experts: the
``Pomerol" is a
rather famous Bordeaux wine). As often emphasized by Landshoff, the
Pomeron
is actually back today in more than one brand: the soft, the hard,
the inclusive and
the exclusive. What are they?
 \begin{itemize}
\item The {\it inclusive} Pomeron, the one controlling total
cross-sections via the
optical theorem,  becomes  {\it soft}  when one deals with soft
interaction physics, in which case it has an intercept $\alpha \sim
1.08$, or
hard, if it refers to Deep Inelastic Scattering (DIS) at very small
$x$. In
the latter case, one talks about the BFKL (from Balitsky, Fadin,
Kuraev, Lipatov)
{\it hard} Pomeron with a much larger intercept ($\alpha \sim 1.15$ --
1.23). Impressive experimental progress has been made  at
HERA\cite{HERA} in
measuring the latter accurately. At the theoretical level,  progress
has
occurred by combining the old Lipatov model with large-$N$ expansion
ideas.
Amusingly, this has led Lipatov, Fadeev, Korchemsky and others to
reformulate the
problem in terms of an Ising ferromagnet\cite{Korchemsky}!

 \item The {\it exclusive} Pomeron, which is measured in diffractive
events.
Again, this can be {\it soft} or {\it hard} depending on the nature
of the events
that are associated with a diffractive trigger. Much exciting
experimental work
has been going on at HERA\cite{HERA} in measuring the so-called
Pomeron structure
function by looking at the Ingelman--Schlein process\cite{IS}, hard
scattering
associated with large rapidity gaps. The data are parametrized  in
terms of the
diffractive structure function $F_2^{D(3)}(x_P,\beta,Q^2)$, a
particular
example of a more general object, the ``fracture
function"\cite{Trentadue}$M_{p,h}^i(z,x,Q^2)$, which describes a
semi-inclusive hard process
initiated by the parton $i$, associated with the detection of a final
hadron $h$ in
the target (here proton) fragmentation region.
\end{itemize}

 For the first time, there have been  reports\cite{HERA} of
measurements of
$M_{p,n}^i(z,x,Q^2)$ which should
be related, in analogy with $M_{p,p}^i$ (the Pomeron case),  to
the pion structure function, an object already measured in the
pion--nucleon
Drell--Yan process. Another interesting piece of news from
DESY\cite{HERA}: the
Pomeron seems to contain hard gluons; actually, maybe it just
consists of a couple
of hard gluons at low $Q^2$, with the rest simply following from
Gribov--Lipatov, Altarelli--Parisi (GLAP)  evolution.
This could be interesting news for the Higgs boson search at hadron
colliders
\cite{Grau}: triggering  on (semi--) diffractive events should increase
 the
gluon-to-quark flux ratio. Since gluon pairs produce Higgs bosons,
while quark pairs
give a two-photon background (the most promising decay channel for a
``light"
Higgs), increasing the gluon/quark ratio increases the
signal/background ratio for
Higgs production.

 \paragraph{\it 3.1.4~$\alpha_s$.}
A new global average was given\cite{Schmell} at the conference:
\beq
\alpha_s(M_Z) = 0.118 \pm 0.003 \; .
\eeq
The good news is that i) the DIS value is no longer ``low"; ii) the
EW/LEP value is
no longer ``high"; iii) lattice calculations are consistent with other
determinations. A compilation of different determinations can be found
in
\cite{Schmell}. In that report one can also find a plot of each value
of $\alpha_s$
against the scale at which the experiment was performed,
and thus ``see" the running of $\alpha_s$.

Before leaving this part of QCD tests, I wish to mention that the
difference
between quark and gluon jets is coming out clearer and clearer from
the data when
one looks at various average properties of the jets e.g. $\langle n
\rangle$,
$\langle p_t  \rangle$, $\langle x  \rangle$. This without mentioning
many other
very fine tests of QCD based on the detailed  analysis of hadronic
final states
\cite{Sterman}.

\paragraph{\it 3.1.5~Top and other heavy quarks.}
Progress has been reported\cite{top} from CDF and D0 at Fermilab on the
top
quark, a crucial component of the SM. The production cross section no
longer
looks too high, while the mass determination $m_t = 175\pm6 {\rm
GeV}/c^2$ is
becoming precise enough to become an important constraint on SM
precision tests
(see Section 3.2).

This is perhaps the right place to mention that the high-$p_t$ jet
cross section
at Fermilab is no longer considered an embarrassment for
QCD\cite{Brock}. New,
improved gluon distributions appear to be consistent with the data,
thus pouring
cold water on some original claims of new physics.

Concerning less-heavy quarks ($c$ and $b$),
theoretical  progress is continuing\cite{heavy} on the heavy-quark
effective
theory (HQET), leading, for instance, to a nice determination of
$|V_{cb}|$ (see
Section 3.2.2).

\paragraph{\it 3.1.6~Heavy ions.}
Last but not least I am coming to the search for signatures of the
quark--gluon
plasma in heavy-ion collisions at CERN. Have we already seen the
first glimpses of
the plasma?  The good news\cite{ions} here comes  from NA50, which
has recently
studied Pb--Pb collisions at $E=158 {\rm GeV}$ per nucleon,
evidentiating a sharp
drop in ${\sigma(J/\psi) \over \sigma_{DY}}$. This ratio had been
proposed long ago
\cite{Satz} as a sensitive probe of  a phase
transition. Unlike the case of $pp$, $pd$, $p-$W, S--U collisions, in
Pb--Pb
collisions the drop cannot  possibly be explained by the nuclear
absorption
cross-section.  From various plots shown  in\cite{ions}, one can
see not only the sharp drop in the Pb--Pb case, but also how such a
drop becomes
more and more pronounced as one looks at ``centrality" bins
corresponding to
increasing $E_T$.

\subsection {Electroweak Interactions}
 Before moving to the ``pi\`ece de
r\'esistance", the status of EW interactions, let me mention a few
other
interesting items that we heard about:

\paragraph{\it 3.2.1~Top quark.}

Besides the already-mentioned top mass and production cross
section, the mixing parameter $|V_{tb}|$ was also
measured with some precision at FermiLab\cite{top}.

\paragraph{\it 3.2.2~Heavy flavours.}

 For the $c$, and even more for the $b$,  HQET
should be reliable. Indeed, good agreement has been found,
particularly in the
process $B \rightarrow D^* l \nu$, which allows  for a good
determination\cite{heavy} of  $cb$ mixing from CLEO data: $|V_{cb}| =
0.0392 \pm 0.0027 (\rm{exp})
 \pm 0.0013 (\rm{th})$, a clear example of what I meant earlier by a
good
matching of theoretical and experimental precision.

\paragraph{\it 3.2.3~Quark masses and mixing}
There have been observations of $B^0_d \bar{B^0_d}$ mixing at LEP, SLD,
CDF,
ARGUS and CLEO giving an interesting
number\cite{mixing} for the mass difference:
\beq
\Delta m_d = 0.464 \pm 0.012 (\rm{stat}) \pm 0.013 (\rm{syst}) \;
ps^{-1},
\eeq
while there are only bounds for $B^0_s \bar{B^0_s}$ mixing from
ALEPH, DELPHI AND
OPAL:
\beq
\Delta m_s > 9.2 \; \rm{ps}^{-1}
\eeq

\paragraph{\it 3.2.4~LEP2.}
I should not pass over the fact that, not long before the start of
this Conference,
we heard that W-pairs had been punctually observed in each one of the
four LEP2
experiments and in a variety of leptonic and hadronic channels. It is
hoped that,
eventually, this will lead to a determination of $M_W$ with a $30$
MeV precision.
At present, CDF and D0 (as well as $\nu N$ DIS data) still provide the
best
determinations, giving a world average:
 \beq
M_W = 80.356 \pm 0.125 \; {\rm GeV}
\eeq

\paragraph{\it 3.2.5~Status of EW interactions.}

Most of the discussion about the status of EW
interactions\cite{Blondel} was based
on the final LEP1 sample  consisting of $1.5\cdot 10^7$ hadronic and
$1.7\cdot
10^6$ leptonic events. The richness of the sample, together with:
\begin{itemize}
\item the precise LEP energy calibration,
\item the normalization of luminosity via Bhabha scattering,
\item the improved understanding of systematics,
\end{itemize}
has led to a spectacular improvement in our knowledge of the EW
parameters.

We can assert that, with LEP1, we have definitely tested the SM at its
very
roots as a renormalizable QFT by achieving sensitivity to QED and
weak radiative
corrections. It is worth while noting that, with the increased
precision of data and
of theoretical calculations, the value of
the fine structure constant itself at the $Z$ mass has become an
 important source of (theoretical) error.   Because of uncertainty in
the running,
$\a(M_Z)$ is only known, at present, at the $0.1\%$  level.

Recalling that $m_t$ was estimated to lie in the right range from EW
precision
data before its actual discovery at Fermilab, we may ask:
Will $m_H$ follow the same
fate? Unfortunately, EW radiative corrections are only
logarithmically sensitive to
$m_H$. Yet, with $m_t$ hopefully known to within $5$ GeV  from
near-future
Fermilab measurements, quantities like ${\rm sin}^2 \theta_W^{eff}$
will have a chance of
distinguishing a rather light Higgs from a rather heavy one.

Let us now go over the status of the main observables:
\begin{itemize}
\item $M_Z = 91.1865 \pm0.0013 \pm 0.0015$.
By its amazing accuracy, this observable  clearly imposes itself as
the third
reference parameter of the standard EW
theory after $\alpha$ and $G_F$. Thanks to the improvements
 mentioned before, $M_Z$ is now
known with a precision of three parts in $10^5$, of which about half
comes from the
beam-energy calibration (which requires, as we heard, not only
knowledge of the
tides but also of the Geneva--Paris TGV schedule!). This is really an
amazing
precision for the
 mass of a particle, especially  considering its width.
\item $\Gamma_Z = 2.4946\pm0.0021\pm0.0017$.
If $M_Z$ defines the EW standard model, $\Gamma_Z$ is crucial in
testing it.
{}From $\Gamma_Z^{inv}$ we can now deduce for the number $N_{\nu}$ of
light SM
neutrinos:
\beq
N_{\nu} = 2.989 \pm 0.012 \; .
\eeq
\item  $R_b$, $R_c$.
Until Warsaw, $R_b$ was the major potential crisis for the SM,
hinting perhaps
towards low-energy supersymmetry, with $R_c$ supporting the picture as
well.
The shaking news we heard is that it may have all been ``much ado
about nothing".
This is mainly due to a new analysis carried out by ALEPH, which
uses a double impact-parameter tag for the $b$ jets (one in each
hemisphere) as well as four more tags to distinguish $b$, $c$ and
$(u,d,s)$ jets.

By measuring five single-tag and 15 double-tag rates, both $R_b$ and
several
efficiencies are estimated. The new ALEPH value:
\beq
R_b = 0.2158 \pm 0.1 \%
\eeq
is on top of the SM prediction... Incidentally, also $R_c$, which was
never that
badly off, is back to the SM value. We see from Blondel's talk that,
while the
old values for $R_b$, $R_c$ excluded the SM at $90 \%$ confidence
level, the new
world averages agree with the SM within $2\sigma$.
We have to realize, however, that a precise determination of $R_b$ is
quite
difficult and not completely assumption-free. Under the
circumstances, my own
conclusion would simply be: if we want to look for a SM crisis, let's
look
elsewhere!
\item $A_b$.
This forward--backward asymmetry parameter appears to be the only
observable
where something like a $3\sigma$ problem can  still be present if LEP
and LSD
data were properly combined.  LSD, with polarized beams, directly
measures  $A_b$,
while at LEP only  $A_e \cdot A_b$ can be extracted. Taking both sets
of
data at face value, one gets:
\beq
A_b = 0.867\pm0.022 \; ,
\eeq
to be compared with the SM's $0.934$. $A_e$ can be converted into a
value
of $0.23165 \pm0.00024$ for
$\sin ^2 \theta_W^{eff}$, which could have
interesting implications.
\end{itemize}

Indeed, from a table of the sensitivity of various observables to
the top and Higgs masses\cite{Blondel},
it is clear  that $\sin ^2 \theta_W^{eff}$ is one of the most
sensitive quantities
to $m_H$ (see\cite{Sirlin} for a recent theoretical appraisal). This
can be seen
again in a plot\cite{Blondel}, where the overlap of all the
constraints defines a
``central region" with $90<m_H<300$. A Higgs at the lower end of this
range would
point in the direction of SUSY, while a heavier one would be more
akin to the
non-supersymmetric SM itself. Unfortunately, this observable is also
quite
sensitive to $\a(M_Z)$, making it important to better determine  the
running of
$\a$.

 \subsection {Gravity}

As already mentioned, gravitational effects are computable in the SM
(here meaning
General Relativity), but are usually tiny.
 Fortunately, through extremely accurate clocks, some of the
very high precision experiments needed to test the theory
have become available\cite{Ramsey}. Time-intervals/frequencies have
becomethe most accurately
known (reproducible) quantities in physics, with one part in
$10^{14}$ becoming
quite standard.  Although there are
not many such tests, they are all quite striking, e.g.:
\begin{itemize}
\item the gravitational red-shift, measured via  an H-maser on a
rocket with a
precision of $0.007\%$,
 \item the $250~\mu$s delay of the  $\mu$-wave signal from the Viking's
passage by the Sun,
 \item the binary pulsar PS1913+16, whose period's rate of
change:
 \beq
\dot{P_b} = -(3.2 \pm 0.6) \times 10^{-12}\; {\rm s} \cdot {\rm
s}^{-1}, \eeq
is in agreement with GR's prediction through energy loss by emission of
gravitational waves (GW). This provides the first (indirect) evidence
for this
crucial prediction of GR. We also know that the speed of GW is the
speed of light
within $0.1\%$.
\item There have  also been claims\cite{Thirring} that the
Lense--Thirring effect has been seen using laser-ranged satellites
\end{itemize}

\renewcommand{\theequation}{4.\arabic{equation}}
\setcounter{equation}{0}
\section {Listening to theory}
 Since the Standard Model works so well, one should perhaps refrain
from asking
questions. Theorists, however, are well known for never being happy.
Let us consider some  of their typical  questions:
 \begin{itemize}
\item How come the lightest ``observed" (inverted commas referring to
quarks and gluons) fundamental particles have $J = {1 \over 2} , 1$?
\item Why are
their mutual interactions described by a gauge theory?
\item Why are  fermions
completely chiral with respect to $G_{gauge}$?
 \item Why is there no light
$\nu_{R}$? \end{itemize} For these typical questions, theorists
provide typical
answers: \begin{itemize}
\item There is a basic difference between scalar, fermion, and gauge
boson masses:
\begin{itemize}
\item gauge boson masses are never allowed (by gauge invariance),
\item fermion masses are not allowed if the fermions are completely
chiral w.r.t.
$G_{gauge}$,
\item bosonic masses are always allowed.
\end{itemize}
\item Even when fermion masses are allowed (say in QCD), there is
an approximate, global chiral symmetry protecting their small value
from
 radiative corrections. Following 'tHooft, we express this by saying
that
fermion masses are ``naturally" small (in the above-mentioned precise
technical
sense for the word ``natural").
\end{itemize}

In order to better understand  whether these are indeed good
answers to the
previous questions, let me digress for a moment on a shift in our
attitude
towards QFT, which occurred around 1980.
The old attitude was that $G_{gauge}$ and $R$ {\it define} a
fundamental
theory, which, through  renormalization, is made finite
and valid at all scales in the limit of infinite cut-off. There is a
price to
pay for the infinities: any hope of  computing a certain number of
physical
constants is lost forever. Such constants must be taken from
experiments (e.g.
the three coupling constants of the SM). This still leaves a lot of
room for
predictivity, as we saw in the previous section.

The more modern attitude consists in saying, more modestly, that QFT is
just
a low-energy {\it effective} theory, while the true
fundamental theory  is a {\it finite} theory endowed with a physical
{\it finite} cut-off $\Lambda$ above which  QFT is no longer valid.
Superstrings today appear as an ``existence proof" for such an ultimate
theory (see Section 8).
In the more fundamental, finite theory,  whatever could
happen did happen, for instance
 all particles that could get a mass (e.g. the $\nu_{R}$) got it.

Starting from this attitude, and invoking the most general
principles of quantum mechanics and special relativity, one can
convincingly argue
\cite{Weinberg} that the {\it effective} theory well below $\Lambda$
{\it must} be
a renormalizable QFT whose Lagrangian
 can be expanded in  powers of $1/\Lambda$ as:
\beq
L^{tree}_{eff} = \Sigma c_n \Lambda^{4-n} O_n \;,
\eeq
\beq
L^{loop}_{eff} \sim {\rm QFT \; loops \; with \; UV \; cutoff}\;
\Lambda
\eeq

The operators $O_n$  have dimensions $mass^n$ and, following a
language borrowed
from statistical mechanics, can be classified as
follows: \begin{itemize}

\item Relevant, for $n<4$,

\item Marginal, for $n=4$,

\item Irrelevant, for $n>4$,
\end{itemize}
where the names come from their relative importance at low energies.
Examples of each kind are:

\begin{itemize}
\item Relevant:  bosonic and fermionic masses. These are
so relevant that, if masses are unprotected and become $O(\Lambda)$
after
radiative corrections, the corresponding particle decouples at low
energy.This explains why we are left with chiral fermions in the
low-energy
domain.
\item Marginal: kinetic terms and gauge couplings. In the absence of
the relevant terms I
just mentioned,  they dominate at low-energy. This ``explains" why the
low energy theory is a chiral gauge theory!
\item Irrelevant: typically operators of dimension $5$ or $6$.
Although suppressed by one or two inverse powers of $E/\Lambda$, such
terms, if present, can show up by contributing to processes that are
otherwise forbidden by relevant and marginal operators (``rare"
processes such as
proton decay or FCNC). Indeed, various global symmetries ($B$, $L_i$
for
each family) are exactly conserved in the SM if one neglects
irrelevant operators.
\end{itemize}

So far so good\dots, except for some puzzles:
\begin{itemize}
\item Some perfectly allowed marginal operators have not been ``seen",
in particular the famous $\Theta$-angle term:
\beq
{\it L}_{\Theta} \sim \Theta_{QCD} \epsilon^{\mu\nu\rho\sigma}
F_{\mu\nu}
F_{\rho\sigma}
\eeq
which contributes to the (not yet observed) electric dipole moment of
the
neutron.
\item  Some relevant operators have not been seen either. The most
distinguished of them is the (in)famous cosmological constant:
\beq
 {\it L}_{cosm} \sim \sqrt{-g} \Lambda_{cosm}\; ,
\eeq
which, unless infinitesimal ($10^{-120}$ in Planck units!), would not
allow
our present large, almost flat Universe to exist. The smallness
(vanishing?) of
the cosmological constant is perhaps the deepest mystery  in physics
today.
 \end{itemize}

Furthermore, following this new attitude towards QFT, we  arrive at a
striking
conclusion:

\centerline{\bf The standard model is not sufficient!}

 Indeed, for any physical
observable $\it{A}$, the  measured value $\it{A}_{phys}$ can be
estimated
theoretically as:
\beq
\it{A}_{phys} = \it{A}_{tree} + \it{A}_{loop}(\Lambda)\; ,
\eeq
where we have indicated explicitly the dependence of the loop
correction
from the UV cut-off $\Lambda$.
 Since $\Lambda$ now does have  a physical meaning, fine-tuning is
strictly
forbidden. This means rejecting
any ad hoc cancellation between $\it{A}_{tree}$ and
$\it{A}_{loop}(\Lambda)$,
which would result in $\it{A}_{phys} << \it{A}_{loop}(\Lambda)$.

Let us apply this general argument to the Higgs boson mass. For an
elementary scalar particle mass $M$, in the absence of supersymmetry,
we
cannot escape the result:
\beq
M_{loop} \sim g \Lambda
\eeq
which clashes, in our philosophy, with the experimental upper limit
(from
precision tests)
\beq
m_{H,phys}  <  O(1~{\rm TeV})
\eeq
unless $\Lambda$ itself is around the TeV scale. This could mean, for
instance, that the Higgs boson is actually non-elementary if studied at
the
TeV scale (see Lane's talk\cite{Lane}).  In the supersymmetric case,
a special
cancellation occurs in $M_{loop}$, so that $\Lambda$ gets replaced by
the
supersymmetry-breaking scale $M_{SUSY}$. In this case, we expect
supersymmetry
to become manifest around 1 TeV. In any case, new physics is expected
below (or around) the TeV scale!

It is perhaps worth while to make a  digression here about a couple
of ``small" conceptual problems within the minimal supersymmetric
standard
model (MSSM). The first has to do with the fact that SUSY
requires (at least) one extra Higgs doublet. The Higgs sector
 of the MSSM is
thus described by the superfields:
\beq
H_1 :(1,2,1) , \;\;\; H_2 :(1,2,-1) \; ,
\eeq
where, as in (\ref{R}), the numbers indicate the representations of
$SU(3)_c \otimes SU(2)_L \otimes U(1)_Y$. The above is obviously a
pair of complex
conjugate representations (since $2 \sim \bar2$). In other words,
a (supersymmetric and
gauge-invariant) mass term $\mu H_1 H_2$ is perfectly allowed, and we
are back to
the puzzle of why such a term is not extremely large (the so-called
$\mu$-problem).
The situation here is only a little better than the one of the Higgs
mass in the
SM. By postulating a new global symmetry, such a mass term can be
made ``naturally"
small (in the technical sense explained earlier).

The second ``problem" is that the automatic (accidental) conservation
of $B$ and
$L_i$ of the SM is generically lost in the supersymmetric Standard
Model (SSM). Some
effort has to be made in order to avoid dangerously large FCNC and,
actually, FCNC
processes at levels close to present experimental limits are almost
unescapable.

\renewcommand{\theequation}{5.\arabic{equation}}
\setcounter{equation}{0}
\section {Hints of new physics?}
 There were a few meager hints of new physics before this Conference.
Well, after Warsaw, none of them is a real hint any longer, perhaps
with
one exception. Let us go quickly through the list:

\begin{itemize}
\item $R_b, R_c, A_{FB}$, see section 3.2.5.
\item $\alpha_s$, see section 3.1.4.
\item Large-$p_t$  events at CDF: see section 3.1.5.
 \item Leptons plus photons plus missing $E_t$ event at
CDF\cite{Brock}. No
confirmation from D0.
 Although the a priori probability of such an
event is small, one should not fall into the trap of confusing a
priori and a
posteriori probabilities:  any   event that has
occurred had zero a priori probability of occurring, of course.
\item Four-jet events seen by ALEPH at LEP1.5\cite{Blondel}.
The  arguments just made apply also to this case. Neither
other LEP experiments, nor new data, have  confirmed ALEPH's original
findings to
this date (the situation has evolved after the Conference: at the time
of
writing,  ALEPH finds four-jet events with similar masses even at
higher
energies, but still without any backing from the other three LEP
experiments).
 \item FCNC. Only upper limits so far,  but sensitivity to
SUSY-induced FCNC is getting closer\cite{Buras}.
 \item Searches for SUSY particles: lower limits are getting tighter
and tighter
\cite{searches}. The controversy continues on Farrar's suggestion of
a light gluino.
  \item Excited quarks, leptons: higher and higher limits again.
\item $\nu$ masses, oscillations:
this remains the only area where evidence for new physics has been
getting better
and better\cite{Suzuki}$^,$\cite{Smirnov} instead of fading away.
There are now three
claims of ``evidence" for oscillations (hence, indirectly, for
non-vanishing
neutrino masses):

\begin{itemize}
\item Solar neutrinos (Homestake, Kamiokande~II, III, Sage and
Gallex)
\item Atmospheric neutrinos ($\nu_{\mu}$ deficit or $\nu_{e}$ excess?)
\item LSND experiment at Los Alamos (hopefully to be confirmed by
Karmen?)
\end{itemize}

The most exciting piece of news, however, is that\cite{Suzuki} Super
Kamiokande
started operating at the beginning of April, and is working very
well! This means
collecting data at a hundred times Kamiokande's rate with checks of
day/night and
seasonal variations. Solar-model-independent tests of the
oscillations thus appear
 feasible and the overall situation should become clear by the time
of the next
ICHEP, in two years from now.
\end{itemize}

 To conclude this section I should remark, in all fairness, that much
of the above
lack of evidence for new physics is not necessarily bad for
supersymmetry since, as
is often emphasized, the SSM is the only known ``gentle" modification
of
the SM, leaving its predictions unchanged below the scale of SUSY
breaking. Yet, one
feels a little frustrated about the lack of any ``direct" evidence
for SUSY and may
start  worrying about a possible overlook of alternative mechanisms for
solving the Higgs-mass problem. After all, before discovering the
Higgs mechanism or
getting the idea of confinement, theory was at a loss for a complete
QFT description
of weak or strong interactions, respectively.
In conclusion, the ``if" appearing in a popular science article last
spring:
``If supersymmetry is true, we are suddenly going to hit a huge new
area of
discovery", still remains an ``if" after Warsaw.

\renewcommand{\theequation}{6.\arabic{equation}}
\setcounter{equation}{0}
\section {What's up in Quantum Field Theory}

\subsection{SUSY-breaking scenarios}

The actual mechanism of SUSY breaking is closely related
 to the possibility of discovering supersymmetry.
Two scenarios of SUSY breaking, distinguished by
who is the carrier of
SUSY breaking,
are in competition at the moment\cite{Ross}.

For quite some time,  people's favourite carrier has been gravity. In
this case, in
order to have an {\it effective} scale of SUSY breaking of about $1$
TeV, one has
to have a much larger scale of SUSY breaking in some ``hidden" sector
that
communicates with us only through gravity. The advantage of this
scheme is that, in
the presence of gravity, SUSY breaking is quite straightforward. The
disadvantage
with a high scale of breaking is that  parameters  have to be
fine-tuned very precisely in order to avoid  problems (typically,
too large FCNC) at low energy.

The new contender is gauge-induced SUSY breaking. Since gauge
interactions are
typically stronger than gravity, this means a {\it lower} scale of
breaking (for
the same $1$ TeV effective scale) and less fine-tuning problems. The
disadvantage
is that the  breaking of global SUSY (i.e. breaking in the absence of
gravity) is
harder (though certainly not impossible) to achieve.

The good news for experimentalists is that the two scenarios are not
only
theoretically  distinct, they also make very different phenomenological
predictions, e.g. on the nature of the lightest SUSY particle (LSP)
which carries
away missing energy and momentum. While in the supergravity scenario
the LSP is
typically the neutralino, in the gauge-mediated scenario it is the
goldstino, the
massless particle that originates from the SUSY analog of Goldstone's
theorem
(the goldstino actually picks up a small mass from gravity and
becomes part of
the massive  gravitino, in the supersymmetric analogue of the Higgs
phenomenon,
but this is not very important phenomenologically).

\subsection{Understanding non-perturbative SUSY dynamics}

Turning now to a more theoretical development, I cannot avoid
mentioning that
recent investigations of
$N=2$ and $N=1$ SUSY gauge theories
have shed entirely new light on non-perturbative phenomena such as
confinement and chiral symmetry breaking. I will illustrate the kind
of progress
that has been made by referring to the already classic paper by
Seiberg and Witten
\cite{SW}
in which a complete (non-perturbative) understanding of $N=2$ SUSY
gauge theories
was achieved. Consider, for simplicity, just the case of an $SU(2)$,
$N=2$ gauge
theory without extra matter multiplets.

This theory has  infinitely many  ground states corresponding to
expectation values of its complex scalar field. Two of these ground
states
stand out: they are characterized by the presence of massless
magnetic monopoles
in their particle spectrum. Upon perturbing the system away from the
$N=2$ limit
(while preserving $N=1$ SUSY), only these two special vacua survive
and therein
monopoles ``condense".  A magnetic Higgs phenomenon thus occurs, in
which electric
charges become confined by  the {\it dual} of the Meissner effect in
Type2
superconductors. In the latter case, an ordinary Higgs phenomenon (due
to
Cooper-pair condensation) spontaneously breaks the gauge symmetry and
confines
magnetic-flux lines into thin tubes. The above results, which can be
proved thanks
to the properties of $N=2$ SUSY, represent a striking confirmation of
the scenario
advocated quite a while ago\cite{tHooft} by 't Hooft and others
for ordinary QCD confinement.

The $N=1$ case has been tackled by Seiberg and
collaborators\cite{Seiberg},extending previous work done in the
eighties. SUSY QCD with gauge group
$SU(N_C)$ has been studied for a long time for a sufficiently small
number $N_F$ of
quark/squark flavours. Seiberg et al.\cite{Seiberg} have investigated
the case in
which $N_F > N_C +2$ and found a correspondence between a strongly
coupled electric
theory at  $N_C +2 < N_F  < 3/2 N_C$  and a weakly coupled magnetic
one at  $3 N_C
< N_F$. Confinement in the electric theory is understood again as a
Higgs phenomenon
 in the dual (magnetic) description.
Finally, for $3/2 N_C < N_F <  3 N_C$,  one predicts a  new
interacting non-Abelian
Coulomb phase. The possibility of extending (some of) these
results to actual (i.e. non-supersymmetric) QCD is still being
explored.

\subsection{Lattice gauge theory}
Turning to lattice gauge theories, I should mention  the following
recent achievements\cite{lattice}:
\begin{itemize}
\item Improvement in the determination of $\alpha_s$, in agreement
now with other
determinations, as discussed in section 3.1.4.
\item Mass of the lightest $0^{++}$ glueball at around $1600$ MeV,
i.e. in the
right range for the candidate discussed under 3.1.1.
 \item
Computation of $|V_{ub}|$, an area where  HQET is not very useful.
\item A new computation of $f_B$ and of $f_{B_s}/f_{B_d}$ giving:
\beq
f_B = 175 \pm 25 {\rm MeV} \; ,
\eeq
and
\beq
f_{B_s}/f_{B_d} = 1.15(5) \; .
\eeq
\item Determination of various $B$-parameters, such as $B_{B_d}$ and
$B_{B_s}/B_{B_d}$.
\end{itemize}
On a more theoretical level:
\begin{itemize}
\item Non-perturbative matching of lattice and $\overline{MS}$
operators,
which is crucial in order to relate quantities ``measured" on the
lattice to those
measured in real experiments.
\end{itemize}

\renewcommand{\theequation}{7.\arabic{equation}}
\setcounter{equation}{0}
\section {What's up in astroparticle physics/cosmology?}

A few subjects would be worth discussing in detail.
 Because of lack of time, I will limit myself to a few quick comments
and refer you
to \cite{Astro} for details:
 \begin{itemize}
\item CMB anisotropies and cosmological parameters.

Data on the Cosmic Microwave Background anisotropy and on large-scale
stucture
 are improving fast and will soon tell us much about the value of basic
cosmological parameters such as $H_0$, the present value of Hubble's
constant.
 \item Microlensing and dark matter

We heard that not many more microlensing candidates have been found
recently.
Yet, this remains one of our main windows on the ``Dark side of the
Universe" as
M. Spiro\cite{Astro} put it. The need for dark matter is still there
today, and
certainly forces us to think that the SM is not the end of physics\dots

\item Gravitational waves: sources and detectors

As already mentioned, we have so far only indirect evidence for the
existence of
GW, a robust prediction of GR. The progress in the field of direct
detection,
either by laser interferometers or by cryogenic resonating antennas,
is slow but
constant. According to experts, there is a definite chance that GW
will be seen by
the turn of the century.
\end{itemize}

\renewcommand{\theequation}{8.\arabic{equation}}
\setcounter{equation}{0}
\section {What's up in string theory?}

In order to justify my spending some time on this esoteric subject,
let me quote
a sentence from the 12 July  issue of the {\it Wall Street Journal}
(this is not a
typo!):

``We have the beginning of a new theory of fundamental physics
--string theory--
whose full elucidation could be as revolutionary as the discovery of
quantum
mechanics or relativity". Now fasten your seat belts!

The main news in this field are related to novel uses of the
duality symmetries and to the so-called {\it D}-Strings and {\it
D}-branes. However,
before giving you an idea of what it's all about, I have to recall
three  basic
``miracles" of string
theory  (see e.g.\cite{GV} for a more detailed account).

 The {\it first}  miracle is that string theory, unlike field theory,
inherits from
quantization a fundamental length parameter $\lambda_s$.  Its
inverse,$M_s \sim \lambda_s^{-1}$, plays the role of the finite UV
cut-off $\Lambda$
discussed in Section 4. String theory, being thus free of UV
divergences, not only
goes over to an effective QFT at energies well  below  $M_s$, but also
pretends to make sense at energies {\it above} it.

 The {\it second}, equally crucial, miracle is that the massless
spectrum of
string theory contains states of all angular momenta up to $2\hbar$.
This
implies, by standard arguments\cite{Weinberg}, that gauge and
gravitational
interactions are automatically present, without having postulated
either a gauge
 or a general covariance principle! Furthermore, gauge and
gravitational couplings
are unified at the string scale, i.e. Coulomb and Newton forces are
automatically of
the same order at the scale $M_s$. This fixes the string scale to be
close to (perhaps one order of magnitude below) the Planck mass.
Summarizing, we
potentially have, in string theory, a finite unified theory of all
interactions.

A {\it third} outstanding property is that QFT's  unphysical,
uncalculable, unmeasurable bare parameters  become,  in string theory,
the
fundamental constants of Nature. Their values are not arbitrary,
God-given numbers,
they are  related to VEVs of fields. Unless they are as such
dynamically
fixed,  there will be  unacceptable
violations of the equivalence principle through new scalar-mediated
long-range
forces.

After this telegraphic introduction, let me  recall what
$T$-duality is. It has been known for some time that closed strings
moving in a
space-time containing extra compact dimensions (which is a typical
situation in
string theory) tend to get\dots ``confused". In particular, they
cannot distinguish
whether the extra dimension is a  circle  of radius $R$ or one of
radius
$\lambda_s^2/R$. This property, known as $T$-duality, has to do with
the interchange of
{\it quantized} momentum (as usual, for a circle, in units ${\hbar
\over R}$) with
the {\it classical} winding of a closed string aroud the circle (the
energy
associated with $m$ windings is proportional, through the string
tension, to
$mR$). Note the {\it quantum} nature of the symmetry: there was no
symmetry before
quantization, since, classically, momentum in a compact direction is
still
continuous.

 At first sight only closed strings can
exhibit $T$-duality, since neither points nor open strings can have
non-trivial
winding. However, as noticed recently by Polchinski and others
\cite{Polchinski}, even open strings do exibit $T$-duality, albeit of
a new kind: a
{\it conventional} open string moving on a circle of radius $R$ is
dual (equivalent)
to a {\it different} kind of open string moving on a circle of radius
$\lambda_s^2/R$.
While the usual open string satisfies  so-called Neumann boundary
conditions (vanishing normal derivative at the end points), the strings
of
the dual theory satisfy Dirichlet boundary conditions. The ends of
these {\it D}-strings ({\it D} for Dirichlet) can only move on a
surface, defined
by fixing some  coordinates, which is called a
 {\it D}-membrane,  {\it D}-brane for short.
{\it D}-branes can also be seen as solitonic solutions of the low
energy field
equations.

 Fine, but what do we get from all this game? At least two applications
have emerged:
\begin{itemize}
\item Certain {\it D}-branes  represent (limiting cases of) black
holes. Now,
the origin of the famous Beckenstein--Hawking formula for the entropy
of a black
hole:
 \beq
S_{bh} = {A \over 4 l_P^2} \; ,
\eeq
where $A$ is the area of the event horizon and $l_P$ is the Planck
length,
has long been a mystery. The exponential of the entropy should be
related to the number of {\it microscopic} quantum states, giving a
definite set of
{\it macroscopic} quantum numbers, i.e. mass, charge, angular
momentum of the  black
hole. So far nobody had been able to identify such microsopic states,
let alone
counting them.

{\it D}-branes were shown to provide explicit examples of how that
counting works.
This could mean making progress in the near future on the famous
information
paradox of black holes the (apparent?) loss of quantum coherence when a
pure
quantum state collapses into a (thermically radiating) black hole,
which eventually
evaporates completely. Steven Hawking, who is the most convinced
advocate of loss
of coherence, looked nervous about the {\it D}-brane breakthrough  at
a talk he gave at
Oxford last June...

\item Until recently, there were five
distinct --but equally consistent-- superstring theories, which looked
to be
completely disconnected from one another. They were (and still are)
going under the
 names of :
\bea
&&  Type I,\; Type IIA ,\; Type  IIB ,\;\nonumber\\&&  Het(SO(32)),\;
Het(E_8 \otimes
E_8)\; .\nonumber \\
\eea

It has become increasingly clear that these theories are actually
related by
duality symmetries. Some of these are of  the $T$-type mentioned above,
others, going under the name of $S$-duality, relate the
strong-coupling limit of
one theory to the weak-coupling limit of another\cite{SF}. This is
similar to the
case of the electric--magnetic duality of Maxwell's theory in the
presence of both
electric  and magnetic charges (monopoles) satisfying Dirac's
quantization
condition:
\beq
q \cdot m = 2 \pi n \; .
\eeq

{\it D}-branes have also played a very useful role in establishing
these links.
As a result, the long-forgotten 11-dimensional supergravity (with a
compact
11th dimension)  has made an impressive come-back, reinterpreted as the
strong-coupling low-energy limit of some 10-dimensional string
theories. It is
actually believed that a new 11-dimensional ``$M$ theory" should
exist, which would
be the ``mother" of all known string theories according to a
remarkable ``genealogical'' tree\cite{SF}.
 \end{itemize}

\renewcommand{\theequation}{1.\arabic{equation}}
\setcounter{equation}{0}
\section {Conclusions}
\begin{itemize}
\item The standard model's health, already excellent, keeps
improving, as it gets
rid of little colds ($R_b$) or of small head--aches (4jet events?).
However:
\item There will be no lack of very interesting new data in the years
to come, and
we are all eagerly waiting for both the expected and (even more so)
the unexpected.
\item Belief in SUSY appears unshakable in the theory community, and
even dangerously contagious for experimentalists. Is it instead
something
completely unthought of that keeps $m_H$ small?
 \item Particle physics has grown to finally
encompass accelerator and non-accelerator experimental physics in an
overall common
effort.
 \item Similarly, understanding the mysteries of Astrophysics,
Cosmology, and Quantum Gravity has become an integral part of our
theoretical
endeavour.
\end{itemize}

And indeed I wish to conclude my talk with a plea. If our field is to
keep
thriving, we cannot afford the luxury of ignoring {\it any} relevant
scientific
input, wherever it may come from, LEP , HERA, COBE, LIGO\dots or a
conceptual
problem in quantum gravity. We have to work hand in hand, theorists and
experimentalists, accelerator and astro physicists, stressing to
ourselves --and to
the public opinion, whose moral and material support we seek--  the

\begin{center}
 {\bf BASIC UNITY OF FUNDAMENTAL PHYSICS}
\end{center}

\noindent where, to avoid any misunderstanding, I should stress that
the word
``fundamental" (like the word ``natural" earlier on) is to be
understood
\begin{center}
{\it in a technical sense},
\end{center}
that would take me too long to explain.

\end{document}